\documentstyle[12pt,aasms4]{article}

\lefthead{Sil'chenko}
\righthead{Young nuclei in the lenticular galaxies. I.}

\begin{document}

\title{Young stellar nuclei in the lenticular galaxies. I. NGC 1023
 and NGC 7332.}

\author{O. K. Sil'chenko\altaffilmark{1}}
\affil{Sternberg Astronomical Institute, Moscow, 119899 Russia\\
       Isaac Newton Institute, Chile, Moscow Branch\\
     Electronic mail: olga@sai.msu.su}

\altaffiltext{1}{Guest Investigator of the RGO Astronomy Data Centre}

\begin{abstract}

As a result of bidimensional spectroscopy of the central parts
of two nearby lenticular galaxies, NGC~1023 and NGC~7332,
undertaken with the Multi-Pupil Field Spectrograph of the
6m telescope of the Special Astrophysical Observatory,
their chemically decoupled stellar nuclei are found to be
substantially younger than the surrounding bulges: the mean
age of the nuclear stellar populations is 7 billion years
in NGC~1023 and $2.5 \pm 0.5$ billion years in NGC~7332.
The morphological analysis undertaken by Seifert \& Scorza (1996)
for NGC~7332 and by us for NGC~1023 has revealed a existence of
separate circumnuclear stellar disks with the radius of 80 pc
in NGC~1023 and of 400 pc in NGC~7332; probably, the intermediate-age
stellar populations are related to these structures.

\end{abstract}


\section{Introduction}

Early-type galaxies, both ellipticals and lenticulars, are usually
thought to be very old stellar systems (see the classic works of
\cite{t68,t72,t78,f73}). Their red integrated colors are consistent
with the mean stellar population age almost as large as that of the
Universe, $T \geq 15$ billion years. Meantime, early-type
galaxies with prominent Balmer absorption lines in their integrated
nuclear spectra indicating a mean age of a few billion years
were observed long ago. Historically, they were found in an
intermediate-redshift cluster and were firstly treated as strongly
evolving objects (\cite{drg83}). Later, the similar population of
galaxies was noted in the rather nearby Coma cluster, and therefore
the phenomenon of "E+A" or "K+A" galaxies was related to the
environment influence (\cite{caletal}). But Zabludoff et al.
(\cite{zetal}) have tested the latter hypothesis using a
homogeneous sample of nearby galaxies and have found that
"{\it $\sim 75$\%\ of nearby "E+A"s lie in the field, well outside
of clusters and rich groups of galaxies}". An overall fraction
of "E+A" galaxies selected by Zabludoff et al. (\cite{zetal})
in LCR Survey is low, only 0.2\%\ of the total list of galaxies;
but they have applied a very extreme selection criterion,
$<\mbox{H}>\,>\,5.5$~\AA. Meantime evolutionary synthesis models
(e. g. of Worthey \cite{worth94}) imply that even if having
$EW(H \beta)\, > \, 2$~\AA\ the stellar population may have
the mean (luminosity-weighted) age less than 5 billion years.
Zabludoff et al. (\cite{zetal}) noted that diminishing their
$<\mbox{H}>$ low limit only to 4.5~\AA\ increases the sample
of "E+A" galaxies by a factor of 3. So we must admit that
a substantial fraction of relatively young stellar
nuclei may be an intrinsic property of the population of present-day
early-type galaxies. Just this conclusion have we made several
years ago, when among $\sim 30$ very nearby lenticular galaxies
about 10 hosts of the nuclei of 5 billion years old or even
younger have been found (\cite{me93a}).

The next question is a question of discreteness. Fisher et al.
(\cite{fish96}) in their detailed investigation of a sample of
nearby lenticular galaxies have also made the conclusion that
the nuclei are on average younger by several billion years than
the bulges; but they prefer to discuss a smooth age increasing
outward, a kind of smooth radial gradient which is characteristic
for galaxy formation by a dissipative collapse. Meantime there
may exist another interpretation: a decoupled nucleus. Seifert \&\
Scorza (\cite{ss96}) have found separate circumnuclear stellar
disks in 7 lenticular galaxies -- in the half of all galaxies
considered by them. We (\cite{we92}) have found chemically decoupled,
metal-rich stellar cores in the lenticular galaxies NGC~1023
and NGC~7332. If the chemically decoupled nuclei are products
of secondary star formation bursts in the centers of early-type
galaxies, a difference of the mean stellar age between the nuclei
and the main galactic bodies must exist: the nuclei must be younger.
The present paper is just devoted to searching for the age
difference between the chemically decoupled nuclei and the surrounding
bulges in the nearby lenticular galaxies NGC~1023 and NGC~7332.

\begin{table}
\caption[ ] {Global parameters of the galaxies}
\begin{flushleft}
\begin{tabular}{lcc}
\hline\noalign{\smallskip}
NGC & 1023 & 7332 \\
\hline\noalign{\smallskip}
Type (NED) & SB(rs)$0^-$ & S0pec  \\
$R_{25}$, kpc (LEDA) & 13.0 & 10.2 \\
$B_T^0$ (LEDA) & 9.55 & 11.50  \\
$M_B$ (LEDA) & --20.46 & --19.93 \\
$B-V$ (RC3) & 0.93 & 0.87 \\
$U-B$ (RC3) & 0.50 & 0.38 \\
 $V_r(radio) $ (LEDA) & 637 $km\cdot s^{-1}$ & 1300 $km\cdot s^{-1}$  \\
Distance, Mpc (LEDA, $H_0$=75 $km\cdot s^{-1}\cdot Mpc^{-1}$) & 11
    & 19.3 \\
Inclination (LEDA) & $90^o$ & $90^o$  \\
{\it PA}$_{phot}$ (LEDA) & $87^\circ$ & $155^\circ$  \\
\hline
\end{tabular}
\end{flushleft}
\end{table}

The global parameters of the galaxies under consideration are
given in Table~1. These galaxies possess some large-scale
peculiarities which may be connected to past merger events.
NGC~1023 is rich of neutral hydrogen, and its distribution and
velocities resemble those of an accreted material (\cite{sancisi}).
However, the global star formation is absent in this galaxy
(\cite{pe93}), and its integrated color is extremely red
(see Table~1), as red as that of a luminous elliptical galaxy.
A neutral hydrogen content of NGC~7332 is low enough, though
non-zero (\cite{knapp}); but it possesses a counter-rotating
disk of ionized gas (\cite{plboul,fish94}).
Otherwise the galaxies look quite normal, moderate-luminous
lenticulars, even non-LINERs. We report our observations and
other data which we use in Section~2. The radial variations of
stellar population age are analysed in Section~3, and the
appearance of inner disks probably related with the age- and chemically
decoupled cores is discussed in Section~4. In Section~5 the two-dimensional
stellar velocity fields obtained with the Multi-Pupil Field Spectrograph
at the 6m telescope are presented. Section~6 gives our conclusions.

\section{Observations}

We have observed NGC~1023 and NGC~7332 in 1996--1997 by using the
Multi-Pupil Field Spectrograph (MPFS, \cite{afetal90,afman}) of the 6m
telescope of the Special Astrophysical Observatory, Nizhnij Arkhyz,
Russia. The journal of the observations is given in Table~2.

\begin{table}
\caption[ ] {2D spectroscopy of NGC~1023 and NGC~7332}
\begin{flushleft}
\begin{tabular}{lllllcc}
\hline\noalign{\smallskip}
Date & Galaxy & Exposure & Configuration & Field
& PA of long side & Seeing \\
\hline\noalign{\smallskip}
15/16.08.96 & NGC~7332 & 40 min & MPFS+CCD $520\times580$ &
$10\arcsec\times16\arcsec$  & $72\arcdeg$ & $2\farcs 4 $ \\
9/10.10.96 & NGC~1023 & 60 min & MPFS+CCD $1040\times1160$ &
$11\arcsec\times21\arcsec$ & $122\arcdeg$ & $1\farcs 6$ \\
31.10/1.11.97 & NGC~7332 & 60 min & MPFS+CCD $520\times580$ &
$10\arcsec\times16\arcsec$ & $167\arcdeg$ & $2\farcs 0$ \\
\hline
\end{tabular}
\end{flushleft}
\end{table}

MPFS, being the second (after CFHT TIGER system, see \cite{bacetal})
realisation of G. Courtes' (\cite{courtes}) concept of spatial sampling
of extended sources by means of a microlens array, is
in active operation at the 6-meter telescope since 1989. Instruments of
this type are providing sufficient gain in investigations of nebulae and
galaxies with respect to a classical long-slit spectroscopy due to
complete coverage of studied area, independence of spectral
resolution on spatial resolution, absence of slit losses and of
the overall problem of object matching.

A set of enlargers, which project the object onto the lens array,
provide spatial sampling according to a
seeing value --- from $\approx$0.3\arcsec\ to $\approx$1.6\arcsec\
per lens. Sizes of the used array varied between 8$\times$11
and 8$\times$16 elements. For this study we have selected the size of
a spatial element $1\farcs 3 \times 1\farcs 3$. The configurations
used in 1996--1997 resulted in 128 spectra per one exposure
for NGC~1023 ($8\times 16$ elements) and in 95 spectra per one
exposure for NGC~7332 ($8 \times 12$ elements). The spectral
range was 4600--5450~\AA\ for NGC~1023 and 4800--5400~\AA\ for NGC~7332
under the spectral resolution of 4--6~\AA\ (dispersion of 1.6~\AA\
per pixel). It includes several strong absorption lines,
$\mbox{H}_\beta$, MgIb, FeI$\lambda$5270, and FeI$\lambda$5335,
which have been used for diagnostics of the stellar population
properties.
To account accurately for the night sky background we have separately
exposed the blank sky region at 1.5\arcmin -- 2\arcmin\ from the
galaxies with an exposure time of one half of that for the galaxies;
the sky was then (after spectrum extraction and linearization) smoothed
and subtracted.

The hollow-cathode lamps filled by He-Ne-Ar or Ne-Xe-Ar gas mixtures were
exposed before and after each exposure in order to provide
wavelength calibration of spectral data. Integrations of the twilight sky
were carried out for correcting system vignetting and
variations of the transmission by the individual lenses.

The primary reduction -- bias subtraction, flatfielding,
cosmic ray hit removing, extraction of one-dimensional spectra,
wavelength calibration, construction of surface brightness maps --
have been fulfiled by using the software developed in the Special
Astrophysical Observatory (\cite{vlas}). After that the
absorption-line indices $\mbox{H}\beta$, Mgb, Fe5270, and sometimes
Fe5335 have been calculated in the standard Lick system
(\cite{woretal}). We have checked our consistency with the Lick
measurements by observing stars from their list (\cite{woretal})
and by calculating the absorption-line indices for the stars in the same
manner as for the galaxies. The indices calculated for 9 stars are
coincident with the data tabulated in (\cite{woretal}) in average within
0.05~\AA. The exposures for the galaxies have been taken long enough
to provide a signal-to-noise ratios of $\sim$ 80--100 in the nuclei
and $\sim$ 30 near the edges of the frames; the corresponding random
error estimations made in the manner of Cardiel et al. (\cite{cgcg})
range from 0.1~\AA\ in the center to 0.5~\AA\ for the individual
spatial elements at the outermost points. To keep a constant level of
accuracy along the radius, we summed the spectra for the galaxies in
concentric rings centered onto the nuclei and studied the radial
dependencies of the absorption-line indices by comparing them to
synthetic models of old stellar populations of Worthey (\cite{worth94})
and Tantalo et al. (\cite{tantalo}). We estimate the mean accuracy of
our azimuthally-averaged indices as 0.1~\AA. To give an impression
on our data quality, the azimuthally-averaged spectra are displayed
in Fig.~1.

Besides our own spectral data, we have used results of the long-slit
observations of NGC~7332 taken from the La Palma Archive. The galaxy
has been observed with the ISIS, the blue arm, of the 4.2m William
Herschel Telescope on August 3, 1994; it was exposed during 30 minutes
in the spectral range 3900--5500 \AA\ with the grating R300B (the
dispersion of 1.54~\AA\ per pixel matched exactly the ours). A slit
with the width of $1\farcs 0$ was aligned with the minor axis of the
galaxy ($PA=65\arcdeg$). In this case also, the Lick indices
$\mbox{H}\beta$, Mgb, Fe5270, and Fe5335 have been calculated along
the slit with binning of 3 pixel ($1\arcsec$).

To study the morphology of the central part of NGC~1023, we have
used photometric data from the La Palma Archive and from the HST
Archive. The galaxy was observed at the 1m Jacobus Kapteyn Telescope
on La Palma through the BVRI filters with the CCD GEC ($0\farcs 3$
per pixel) on October 24, 1990, and on September 11, 1991, under
moderate seeing conditions, $FWHM_*=1\farcs 8 - 2\farcs 0$. Also it
was exposed with WFPC2 of the HST through the filters F555W and F814W
on January 27, 1996, under the spatial resolution of $0\farcs 1$
(Principal Investigator: S. M. Faber; Program ID: 6099). All the
data have been analysed with the program of V. V. Vlasyuk (SAO RAS)
FITELL.

\section{Ages of the Stellar Nuclei in NGC~1023 and NGC~7332}

To increase signal-to-noise ratios and to derive more precise radial
profiles of the absorption-line indices, we integrate our field
measurements in circular concentric rings centered onto the galactic
nuclei. But the galactic bulges are rather flattened: the isophote
ellipticities in the radius range $3\arcsec \div 8\arcsec$ are nearly
0.2 in NGC~1023 and 0.4 in NGC~7332. Besides, there may be
multicomponent photometric structure, perhaps, a noticeable disk
influence along the major axes in the edge-on lenticular galaxies
under consideration. We would like to be sure that by using the
azimuthally averaged radial profiles of the absorption-line indices
we compare the nuclei to their surrounding bulges and not to some
artificial unphysical units.

For NGC~7332 we can make our verification by comparing our
azimuthally averaged profiles to the linear cross-section along
the minor axis (so to the "pure" bulge). Figure~2 presents the
results of this comparison. First of all, we would like to note
a good agreement between both sets of measurements with the
MPFS: the differences inside $R \approx 5\arcsec$ do not exceed
0.2~\AA\ confirming our estimate of the index accuracy, 0.1~\AA.
Secondly, the long-slit measurements along the minor axis, being
less accurate than the azimuthally averaged data, however, follow
the MPFS profiles rather well; al least there is no systematic
difference between two kinds of profiles. We should only add that
in the external parts of the profiles, $R > 5\arcsec$, where our
data of 1996 and of 1997 are somewhat diverging, the long-slit
measurements confirm the 1997's version, particularly, of the
$\mbox{H}\beta$ and Fe5270 profiles.

For NGC~1023, with its high surface brightness in the center, we
have been able to calculate two-dimensional maps of the absorption-line
indices. The maps for Mgb and Fe5270 slightly smoothed at
$R \geq 2\arcsec$ (by a two-dimensional gaussian with
$\sigma = 1\farcs 0$) are presented in Fig.~3. Surprisingly, they
look different: if the isolines on the Fe5270 map are elongated
nearly as the isophotes and may be attributed to the effect of the
(probable) metal-rich circumnuclear stellar disk, the isolines on
the Mgb map are completely decoupled and may be described as
elongated too but prominently turned.
So the analysis of the two index maps prevents us from a selection
of any particular ellipsoidal trajectories of index averaging.
Taking into account the results of both considerations, for
NGC~7332 and for NGC~1023, we conclude that the use of the
circularly averaged absorption-line index profiles for comparison
of the nuclei to the underlying bulges is the most reliable
as well as the simplest way.

The next Figures are "index-index" diagrams where we compare the
nuclei of NGC~1023 and NGC~7332 to the surrounding bulges. To compare
our measurements to the stellar population models based on summation
(with some weights) of spectra of stars, we must made corrections
for the stellar velocity dispersion in galaxies which broaden absorption
lines and "degrade" a spectral resolution in such way. We have
calculated the corrections by smoothing the spectra of K0-K3 III giants
from the list of (\cite{woretal} which we have observed and by
measuring the absorption-line indices of the smoothed spectra.
We have found that the index H$\beta$ is quite insensitive to the
velocity dispersion when $\sigma_v$ remains to be less than 230 km/s;
as for metal-line indices, we have found that the corrections are:\\
0.1~\AA\ for $\sigma_v$=130 km/s (NGC~7332, \cite{sp97}),\\
0.3~\AA\ for $\sigma_v$=180 km/s (circumnuclear regions of
NGC~1023, \cite{sp97}), and\\
0.4~\AA\ for $\sigma_v$=215 km/s (the nucleus of NGC~1023, LEDA).\\
All the Figures beginning from Fig.~4 contains the corrected indices;
for comparison, the data from Trager et al. (\cite{trager}) for the
nuclei of NGC~1023 and NGC~7332 are plotted too -- thought they are
less accurate than ours, the agreement within their errors is good.

The diagrams (Fe5270, Mgb) are intended to check if the magnesium-to-iron
ratios are solar. We want to check it because the most age-metallicity
diagnostics are calculated for the solar elemental ratios; besides
that, the magnesium-to-iron ratio is important itself because it
characterizes a duration of the main star formation epoch. We know
that luminous elliptical galaxies are mostly magnesium-overabundant
(\cite{worth92}); but as for lenticulars, there were discordant
opinions: we (\cite{me93b}) found that spectroscopic observations
of bright S0 galaxies through a
$4\arcsec \times (1\arcsec \div 2\arcsec)$ aperture revealed mostly
solar magnesium-to-iron ratios while Fisher et al. (\cite{fish96})
found the brightest S0 galaxies to have magnesium-overabundant
nuclei, just as ellipticals. Fig.~4 allows to resolve this problem
with respect to NGC~1023 and NGC~7332. The former has an obviously
magnesium-overabundant nucleus; but starting from $R \approx 1\farcs 3$
all the bulge measurements follow tightly the model sequences for
[Mg/Fe]=0. So we conclude that in the nucleus of NGC~1023
[Mg/Fe]$\approx +0.3$ and in the bulge [Mg/Fe]$\leq +0.1$. As for
NGC~7332, we can surely state that it has a solar magnesium-to-solar
ratio in the nucleus, because we have measured the nuclear indices
very accurately. The bulge measurements follow the [Mg/Fe]=0 model
sequences up to $R\approx 4\arcsec$; after that two measurement sets
diverge: the data of 1996 continue to follow the model sequences
while the data of 1997 show a slight magnesium overabundance. Though
the long-slit measurements support rather the results of 1997 in
this radius range (see Fig.~2), we would prefer to conclude that
the bulge of NGC~7332 demonstrates nearly solar magnesium-to-iron
ratio, just as the nucleus.

It is known that both age and metallicity decreases result in
the bluer color and weaker metal absorption lines in the integrated
spectra of the stellar populations. However there are methods
to disentangle age and metallicity effects. Particularly, comparing
Balmer-line and metal-line indices, one can determine simultaneously
both parameters. The most popular present models which provide a lot
of various absorption-line indices for old stellar populations
($T > 1$ billion years) are the models of Worthey (\cite{worth94});
however they are calculated for [Mg/Fe]=0. Recently some advanced
models have been published; among them we take the results of
Tantalo et al. (\cite{tantalo}) which are expanded to [Mg/Fe]=--0.3
and [Mg/Fe]=+0.3. In NGC~1023 only the star-like nucleus has
[Mg/Fe]$\approx +0.3$, and the rest of the region under consideration
has [Mg/Fe]$\approx 0.0$. So we need both sets of calculations for
the age-metallicity diagnostics.
To determine an age of the star-like nucleus, we must use the Fig.~5a
where the models of Tantalo et al. (\cite{tantalo}) for [Mg/Fe]=+0.3
are plotted. We see that the nucleus itself is $\sim$ 7 billion years
old and has more than solar global metallicity. Fig.~5b which contains
the comparison with the models of Tantalo et al. (\cite{tantalo})
calculated for [Mg/Fe]=0.0 evidences that the ring at $R=1\farcs 3$
is young, nearly 5 billion years old, and has more than solar global
metallicity; the bulge is $\sim$ 15 billion years old and moderately
metal-poor. Fig.~5c presents a comparison to the Worthey's models,
so is valid for the measurements except the nucleus. One can see
once more that the point at $R=1\farcs 3$ confirms
rather young age of the stellar population in the nearest vicinity
of the nucleus, namely,  $T\leq 5$ billion years; more outer bulge,
at $R > 2\arcsec$, is old, $T \sim 17$ billion years. So we conclude
that there exists a compact, marginally resolved stellar structure
in the center of NGC~1023, with the radius of $R \leq 1\farcs 5$,
which is significantly younger than the surrounding bulge. There is
also a modest metallicity difference between the two:
the nuclear+circumnuclear stellar population has [m/H]$\approx +0.3$,
while in the central bulge we see [m/H] from 0.0 to --0.3.

The work of Tantalo et al. (\cite{tantalo}) proposes also a possibility
to quantify differences of stellar population properties basing on
the index differences. A set of three linear equations, connecting
$\Delta$[Mg/Fe], $\Delta \log Z$, and $\Delta \log T$ to the
$\Delta \mbox{Mg}_2$, $\Delta <\mbox{Fe}>$, and $\Delta \mbox{H}\beta$,
is proposed. We apply these equations to the differences between the
nucleus and the bulge or between the nucleus and circumnuclear structure
(the ring at $R=1\farcs 3$) in NGC~1023; the bulge is taken by integrating
the rings at the following values of radius, $r=2\farcs 6$, $3\farcs 9$,
and $5\farcs 2$. Having performed the set of calculations,
we have obtained the following parameter differences:
for the difference "circumnuclear ring -- minus -- nucleus"
$\Delta$[Mg/Fe]=-0.20, $\Delta \log Z$=+0.09, and $\Delta \log T$=-0.19;
and for the difference "bulge -- minus -- nucleus"
$\Delta$[Mg/Fe]=-0.27, $\Delta \log Z$=-0.27, and $\Delta \log T$=+0.26.
It means that the bulge is twice older and twice metal-poorer
than the nucleus; the circumnuclear structure
is almost as metal-rich as the nucleus but is younger by a factor of 1.5.
Interestingly, the Mg/Fe ratios are almost equal in the circumnuclear
structure and in the bulge, but the nucleus is outstanding by its
magnesium overabundance. The bulge has a solar magnesium-to-solar
ratio as one can see from Fig.~4a; then the nucleus has
[Mg/Fe]$\approx +0.3$, and the age estimate for it obtained from the
diagram Fig.~5a is valid.

For NGC~7332 we can use the models with [Mg/Fe]=0 for the nucleus and
for the bulge as well. But there are some other restrictions: the
data of 1997 lack extended $< \mbox{Fe} >$ measurements though their
$\mbox{H}\beta$ indices are more accurate at $R \geq 4\arcsec$ (see
Fig.~2). So we give three variants of "$\mbox{H}\beta$, metal index"
diagrams to disentangle age and metallicity effects in NGC~7332.
Fig.~6a reveals a strong age gradient in the radius range
$0 -5\arcsec$; outside $R \approx 5\arcsec$ the age of the stellar
population remains constant and older than 17 billion years. Figs.~6b
and 6c confirm the age increase by a factor of 6 between $R=0\arcsec$
and $R=5\arcsec$. And all three diagrams show surely that the mean
age of the nuclear stellar population in NGC~7332 is less than 3
billion years. Moreover, as we have three independent high-quality
nucleus measurements (see Fig.~2), we can calculate the H$\beta$
index value very accurately: $\mbox{H}_{\beta}(nuc)=2.20\pm 0.05$~\AA;
(and the metal-line indices also: Mgb=$3.82\pm 0.04$~\AA\ and
$< \mbox{Fe} > = 2.95\pm 0.03$~\AA). It lets a very precise estimate
of the nuclear stellar population age: $T=2.5\pm 0.5$ billion years
with the [m/H]$\approx +0.3 - +0.4$. This age is unexpectedly low for
a regular gas-poor lenticular galaxy. A metallicity drop between the
nucleus and the bulge at $R \geq 5\arcsec$ is also very prominent:
from +0.3 to $-0.5 \div -0.7$, almost an order of magnitude. Therefore,
in NGC~7332 the age- and metallicity-decoupled circumnuclear structure
is extended, with the $R\approx 5\arcsec$ (it is somewhat overestimated
value due to our moderate seeing quality); outside it we see
an unusually metal-poor bulge with the old stellar population.

\section{Morphology of the Central Parts of NGC~1023 and NGC~7332}

There are several photometric studies of NGC~7332 which have been
made recently, with CCD detectors, under rather good seeing conditions
and followed by a sophistic mathematical analysis (\cite{fish94},
Seifert \&\ Scorza \cite{ss96}). The paper of Fisher et al. (1994)
is fully devoted to the dynamics and structure of NGC~7332. The
photometric data obtained with the 3m Lick Telescope have allowed
them to decompose the radial brightness profile into a bulge, an
extended exponential disk, and a third component, "something flat"
in the radius range of $14\arcsec - 24\arcsec$. The kinematical data
have revealed a presence of counterrotating ionized gas, but nothing
unusual in the rotation of the stellar component. Meantime, the
stellar velocity dispersion has a prominent minimum in the center
of the galaxy which cannot be consistent with the dominance of the
hot spheroidal bulge. Seifert \&\ Scorza (\cite{ss96}) have made
two-dimensional decomposition of the surface brightness map of
NGC~7332, together with other lenticular galaxies. They have found
two disks in this galaxy: the inner one has a maximum surface brightness
at $R\approx 3\farcs 5$ and the outer one -- at $R\approx 23\arcsec$.
Between two disks there is a gap: at $R\approx 10\arcsec$ only a
spheroidal component is detected. We would like to note that the
radius ranges for the inner disk photometric incidence and for the
region of the strong age gradient reported in the previous Section
are roughly the same.

NGC~1023 being brighter and larger than NGC~7332 was intensively studied
in the epoch of photographic photometry. Probably, a general impression
that Barbon \&\ Capaccioli (\cite{bc75}) and Gallagher \&\ Hudson
(\cite{gh76}) have made all the possible prevented an investigation of
NGC~1023 with digital detectors. Barbon \&\ Capaccioli (\cite{bc75})
have derived three-component structure of the radial surface brightness
profile: de Vaucouleurs' bulge at $R \leq 0\farcm 5$, an exponential
disk at $R \geq 1\arcmin$ and a lense between them -- a picture very
similar to the structure of NGC~7332 derived by Fisher et al. (1994).
But a two-dimensional analysis a la Seifert and Scorza (\cite{ss96})
was still needed. Figure~7 presents some results of such analysis
which we have undertaken by using the photometric data from the
archives of the HST and La Palma. One can see that the HST data are
more precise (though limited by a smaller radius range), but in
general the agreement between the different telescope and different
passband results exists. High spatial resolution of the HST observations
enables us to detect a very compact distinct structure in the center
of NGC~1023. It manifests itself as a local maximum of ellipticity
at $R\approx 1\arcsec$. Since the fourth cosine Fourier coefficient is
larger than +1\%\ inside $R=1\farcs 5$, we can conclude that this
distinct subsystem is a compact nuclear stellar disk. We have
subtracted pure elliptical surface brightness distribution from
the HST image of NGC~1023; the result is shown in Fig.~8. Even though
we have ascribed the ellipse parameters affected by the disk
presence to our model image, the residual map still demonstrates
a presence of the thin edge-on disk with a radius of $\sim 1\farcs 5$.
The orientation of the nuclear disk, $PA\approx 84\arcdeg - 85\arcdeg$,
is very close to the global line of nodes, $PA_0=86\arcdeg$ (Barbon
\&\ Capaccioli \cite{bc75}), though the bulge itself -- or rather
the lense if it dominates at $R=35\arcsec - 40\arcsec$ -- reveals some
misalignment, by $10\arcdeg - 12\arcdeg$, with respect to the outermost
disk. If we deal in this radius range with some triaxial subsystem
-- NGC~1023 is classified as SB0, -- this misalignment transforms in the
plane of the galaxy into a very prominent turn taking into account
that the galaxy is seen almost edge-on ($i=80\arcdeg$, Barbon \&\
Capaccioli \cite{bc75}). If this lense is a disk-like subsystem, which
is implied by high values of $a4$ between $R=20\arcsec$ and
$R=40\arcsec$, it demonstrates a strange local "warp" scarcely
explicable in the inner region of the early-type galaxy which is
dominated by an oblate spheroid. Detailed kinematical data are needed
to classify this subsystem intermediate between the bulge and the
main disk. But inside $R\approx 14\arcsec$ the structure seems to be
clear: the compact circumnuclear disk limited by $R\approx 1\farcs 5$
is embedded into the pure ellipsoidal bulge. This morphology is quite
consistent with the age trend detected in the previous Section:
the entity with more than solar metallicity and the age younger than 10
billion years appears to be a nuclear stellar disk. The bulge is old
and moderately metal-poor.

\section{Stellar Rotation in the Inner Parts of NGC~1023 and NGC~7332}

The spectral range which has been exposed with the MPFS contains a lot
of strong absorption lines. So it has been a possibility to calculate
a field of stellar velocities by cross-correlating elementary galactic
spectra with spectra of some K-giants which has been observed the same
nights as the galaxies. Two-dimensional line-of-sight velocity maps
obtained are presented in Fig.~9. Both galaxies are seen nearly edge-on.
The stellar velocity field of NGC~1023 (Fig.~9a) can be treated as
cylindric rotation; but the stellar velocity field of NGC~7332 (Fig.~9b)
is not so conventional: a sure local velocity extremum is clearly seen
to the north from the center which may be a sign of a dynamically
decoupled core. In any case, rotational velocities can be derived from
major-axis cross-sections which are simulated from the two-dimensional
maps and presented in Fig.~10. For comparison and to extend the
rotation curves, the long-slit data from \cite{sp97} are also plotted;
to compare them, we have reduced the profiles to the same systemic
velocities: in the case of NGC~1023 the difference of the systemic
velocity is 37 km/s, and in the case of NGC~7332 -- 28 km/s. But the
shapes of the central velocity gradients according to our observations
and to those of \cite{sp97} are very similar. For NGC~1023 (Fig.~10, top)
the agreement is perfect; the combination of the data presented lets
to conclude that the central part of the galaxy inside
$R\approx 3\arcsec$ (it is an upper limit due to finite spatial
resolution) rotates much more rapidly than the rest of the galaxy.
An analogous conclusion may be made with respect to NGC~7332 if
we believe in our MPFS cross-section: the velocity profile from
\cite{sp97} looks somewhat shallower than ours. Since our profile
is symmetric and the profile from \cite{sp97} is clearly asymmetric
and since during long-slit observations there may be difficulties
with slit positioning, perhaps, our data are more reliable. So we
may conclude that the central parts of NGC~1023 and NGC~7332 look
rotating faster than the surrounding bulges and are kinematically
decoupled.

\section{Conclusions}

We have found chemically distinct, in average relatively young stellar
cores in the lenticular galaxies NGC~1023 and NGC~7332. The morphological
analysis undertaken by Seifert \&\ Scorza (\cite{ss96}) for NGC~7332
and here by us for NGC~1023 has implied them to be related to the
separate circumnuclear stellar disks. The similar conclusion about the
secondary formation epoch has been made by de Jong and Davies (\cite{jd})
for embedded disks of some discy ellipticals, when they have found
correlation between H$\beta$ index and the fourth cosine Fourier
coefficient of isophotes. By considering stellar rotation
curves of NGC~1023 and NGC~7332, we see that the central (chemically
decoupled) regions are also distinguished by a fast solid-body rotation,
so they can be also described as dynamically decoupled. Interestingly,
the borders of the regions distinct by the metallicity, by the age, and
by the morphology are approximately the same --
that is, $R=1\farcs 5$, or 80 pc, for NGC~1023, and $R=4\arcsec$, or
400 pc, for NGC~7332. In the case of NGC~7332 we are sure to resolve
the decoupled region, and we can state rapid changes of stellar
characteristics, especially age, along the radius due to the circumnuclear
disk effect, the nucleus being the youngest point. The decoupled
substructure in NGC~1023 though very compact seems to be marginally
resolved too. The difference in magnesium-to-iron ratio between the
nucleus and the ring at $R=1\farcs 3$ looks convincingly: it is consistent
with the age difference of 2 billion years implying that the secondary star
formation burst was short in the nucleus and lasted for several (2--3)
billion years at the periphery of the circumnuclear disk. The
coincidence of chemically, morphologically and dynamically decoupled
central regions was also found in another lenticular galaxy, NGC~4816
(\cite{mehetal}); in this galaxy the circumnuclear disk is also
relatively young, $T \leq 8$ billion years, and very strongly
dynamically decoupled (being counter-rotating). Perhaps, in this relation
lenticular galaxies are similar to ellipticals where the coincidence
of chemically decoupled cores with fast-rotating circumnuclear stellar
disks is frequent (Bender \&\ Surma \cite{bs92},
\cite{sb95,scb95,ffi95}). In early-type spiral galaxies the situation
looks somewhat different: we have found that in M~31 (\cite{silbv98})
and in NGC~4216 and NGC~4501 (\cite{silbv99}) the chemically decoupled
young stellar cores are much more compact than the circumnuclear
stellar disks, although both are present.

\acknowledgements
I thank the astronomers of the Special Astrophysical Observatory
Drs. V. L. Afanasiev, A. N. Burenkov, S. N. Dodonov, V. V. Vlasyuk,
and Mr. Drabek for supporting the observations at the 6m telescope.
I am also grateful to the graduate student of the Moscow University
A. V. Moiseev for the help during the observations and in
preparing the manuscript.
The 6m telescope is operated under the financial support of
Science Department of Russia (registration number 01-43).
During the data analysis I have
used the Lyon-Meudon Extragalactic Database (LEDA) supplied by the
LEDA team at the CRAL-Observatoire de Lyon (France) and the NASA/IPAC
Extragalactic Database (NED) which is operated by the Jet Propulsion
Laboratory, California Institute of Technology, under contract with
the National Aeronautics and Space Administration.
This research has made use of the La Palma Archive. The telescopes
WHT and JKT are operated on the island of La Palma by the Royal
Greenwich Observatory in the Spanish Observatorio del Roque de los
Muchachos of the Instituto de Astrofisica de Canarias.
The work is partly based
on observations made with the NASA/ESA Hubble Space Telescope, obtained
from the data archive at the Space Telescope Science Institute, which is
operated by the Association of Universities for Research in Astronomy,
Inc., under NASA contract NAS 5-2655.  We have also used the software
ADHOC developped at the Marseille Observatory, France. The work
was supported by the grant of the Russian Foundation for Basic
Researches 98-02-16196, by the grant of the President of Russian
Federation for young Russian doctors of sciences 98-15-96029
and by the Russian State Scientific-Technical
Program "Astronomy. Basic Space Researches" (the section "Astronomy").

\newpage

\figcaption{Azimuthally averaged spectra obtained with MFPS for
NGC~1023 ({\it a}) and NGC~7332 ({\it b}). The corresponding radii
are 0\arcsec, $1\farcs 3$, $2\farcs 6$, $3\farcs 9$, $5\farcs 2$,
$6\farcs 5$, and $7\farcs 8$, from top to bottom. The normalizations
are arbitrary}

\figcaption{NGC~7332: Radial profiles of the absorption-line indices
$\mbox{H}\beta$, Mgb, Fe5270, and Fe5335; filled and open circles
present our MPFS data, small triangles and squares are two sides of
the long-slit WHT ISIS cross-section along the minor axis of the
galaxy}

\figcaption{Two-dimensional maps of the central part of NGC~1023:
 {\it a} -- continuum $\lambda$5000, in arbitrary intensity units,
 {\it b} -- the Lick index Fe5270, in 0.01~\AA,
 {\it c} -- the Lick index Mgb, in 0.01~\AA}

\figcaption{The diagrams (Fe5270, Mgb): {\it a} -- for NGC~1023,
{\it b} -- for NGC~7332.
The measurements are azimuthally averaged and taken along the radius
with the step of $1\farcs 3$.
For NGC~7332 the mean nucleus position (averaged over two MPFS sets and
one long-slit spectrum) is also shown with the appropriate error bar.
The nuclear measurements from Trager et al. (1998) are plotted for
comparison with their error bars.
The ages of the Worthey's (1994) models are given in billion years}

\figcaption{The age-diagnostics diagrams for NGC~1023: {\it a} --
H$\beta$ vs $< \mbox{Fe} >$ for [Mg/Fe]=+0.3
(the models of Tantalo et al. 1998), valid for the nucleus,
{\it b} --
H$\beta$ vs $< \mbox{Fe} >$ for [Mg/Fe]=0 (the models of Tantalo et
al. 1998), valid for all the points except the nucleus,
{\it c} --
H$\beta$ vs [MgFe] for [Mg/Fe]=0 (the models of Worthey 1994), valid
for all the points except the nucleus.
The measurements are azimuthally averaged and taken along the radius
with the step of $1\farcs 3$.
The nuclear measurements from Trager et al. (1998) are plotted for
comparison with their error bar.
The ages of the models are given in billion years;
the metallicities for the Worthey's models are +0.50, +0.25, 0.00,
--0.22, --0.50, --1.00,--1.50, --2.00, if one takes the signs from
the right to the left, and for the models of Tantalo et al. they are
+0.4, 0.0, and -0.7}

\figcaption{The age-diagnostics diagrams for NGC~7332: {\it a} --
H$\beta$ vs Mgb, the models of Worthey (1994), {\it b} --
H$\beta$ vs [MgFe], the models of Worthey (1994), {\it c} --
H$\beta$ vs $< \mbox{Fe} >$ for [Mg/Fe]=0, the models of Tantalo et
al. (1998).
The measurements are azimuthally averaged and taken along the radius
with the step of $1\farcs 3$.
The mean nucleus position (averaged over two MPFS sets and
one long-slit spectrum) is also shown with the appropriate error bar.
The nuclear measurements from Trager et al. (1998) are plotted for
comparison with their error bars.
The ages of the models are given in billion years;
the metallicities for the Worthey's models are +0.50, +0.25, 0.00,
--0.22, --0.50, --1.00,--1.50, --2.00, if one takes the signs from
the right to the left, and for the models of Tantalo et al. they are
+0.4, 0.0, and -0.7}

\figcaption{Radial variations of the isophote morphological
characteristics in the center of NGC~1023 according to the HST data
and to the La Palma data}

\figcaption{The residual brightness map of the central part of
NGC~1023 obtained from the HST WFPC2 F814W image by subtracting the
pure ellipsoidal brightness distribution. The sizes of the map are
$23\arcsec \times 21\arcsec$, the orientation is $PA(top)=120\arcdeg$,
the black spot in the center marks the nucleus position, the light
ellipse borders the area where the model has been subtracted}

\figcaption{Stellar isovelocities in the centers of NGC~1023 ({\it a})
and NGC~7332({\it b}). The positions of photometric centers are
shown by the crosses}

\figcaption{Major-axis line-of-sight velocity profiles for the
stellar components in the centers of NGC~1023 ({\it a}) and NGC~7332
({\it b})}

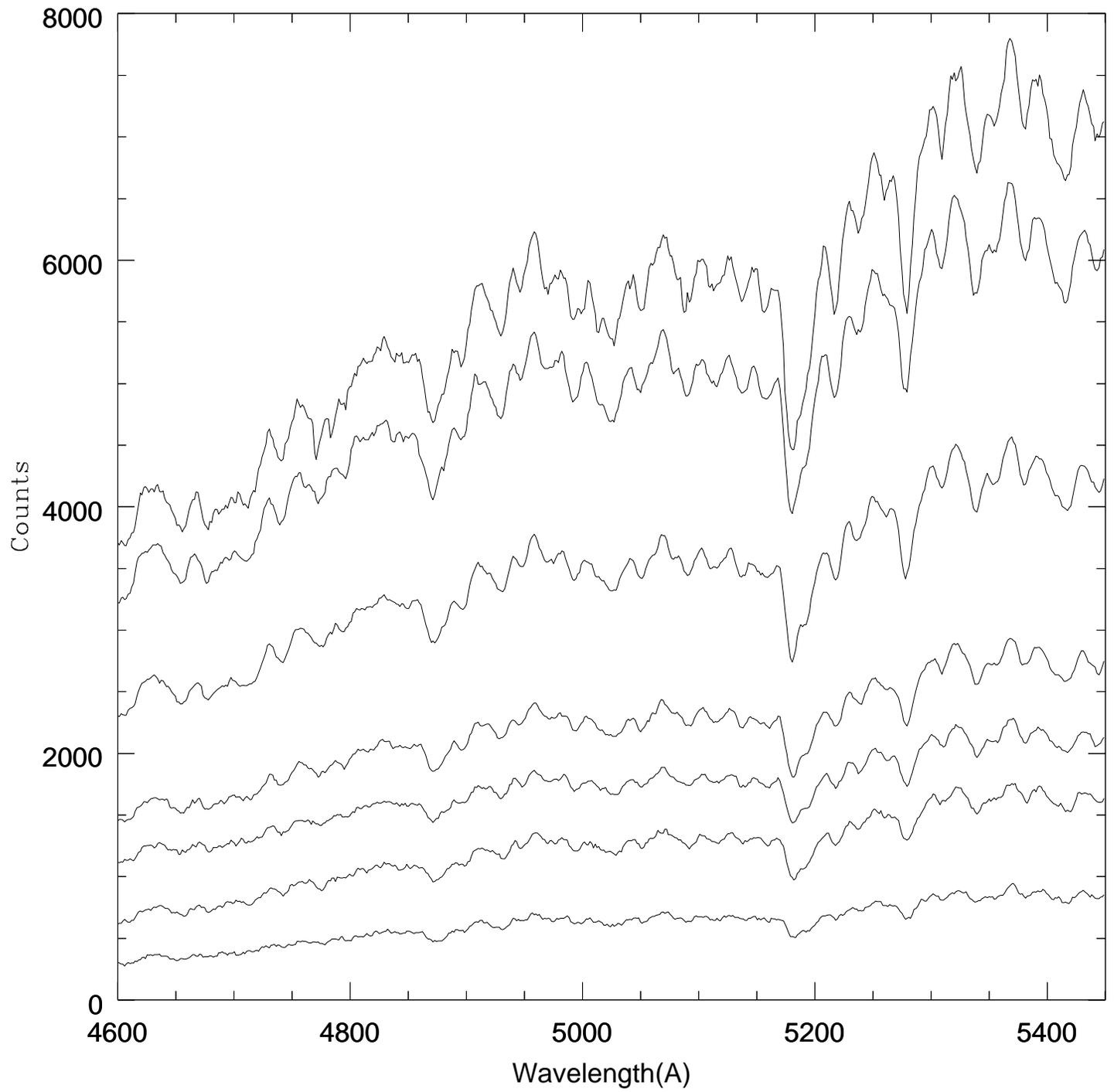
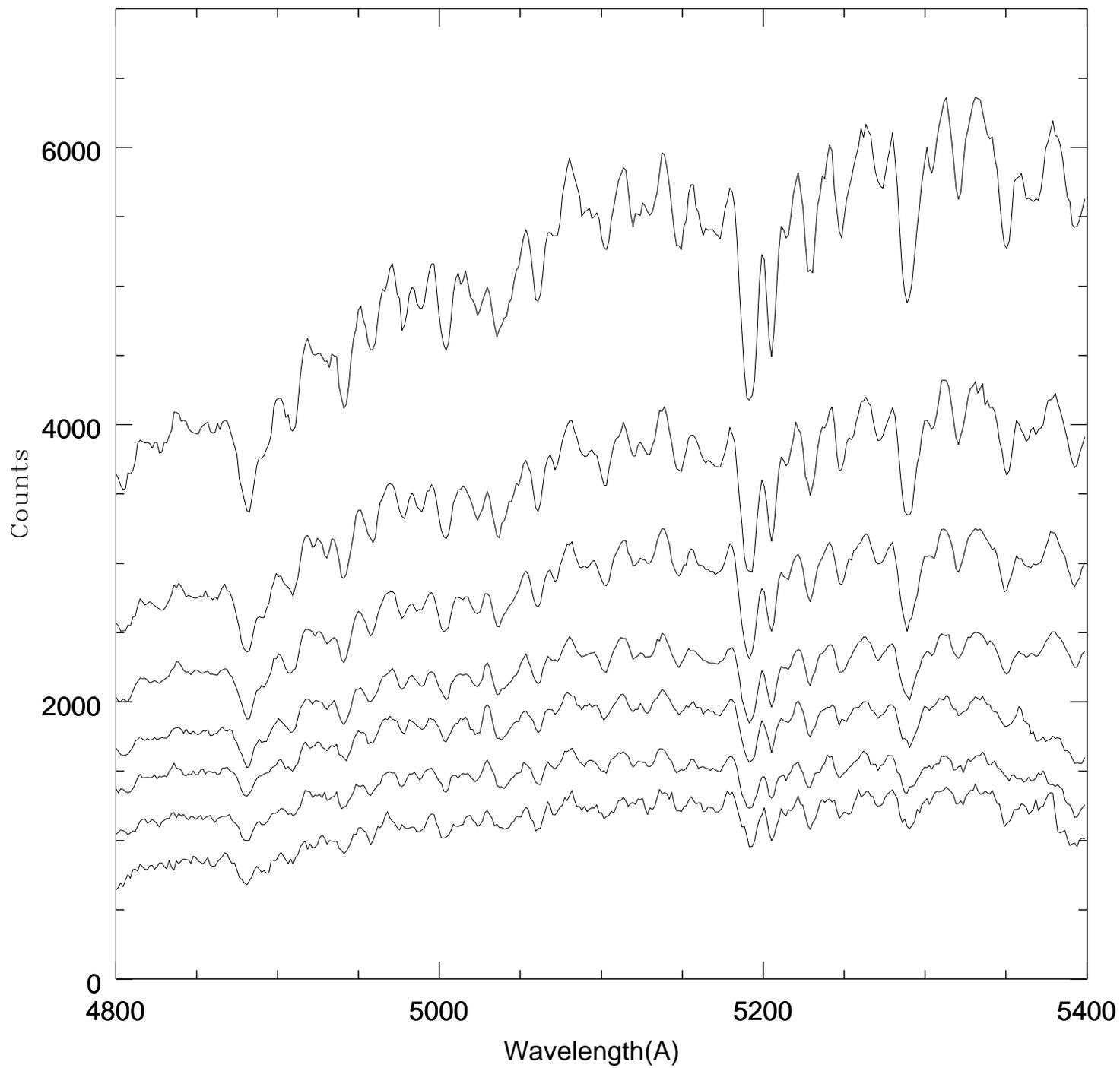
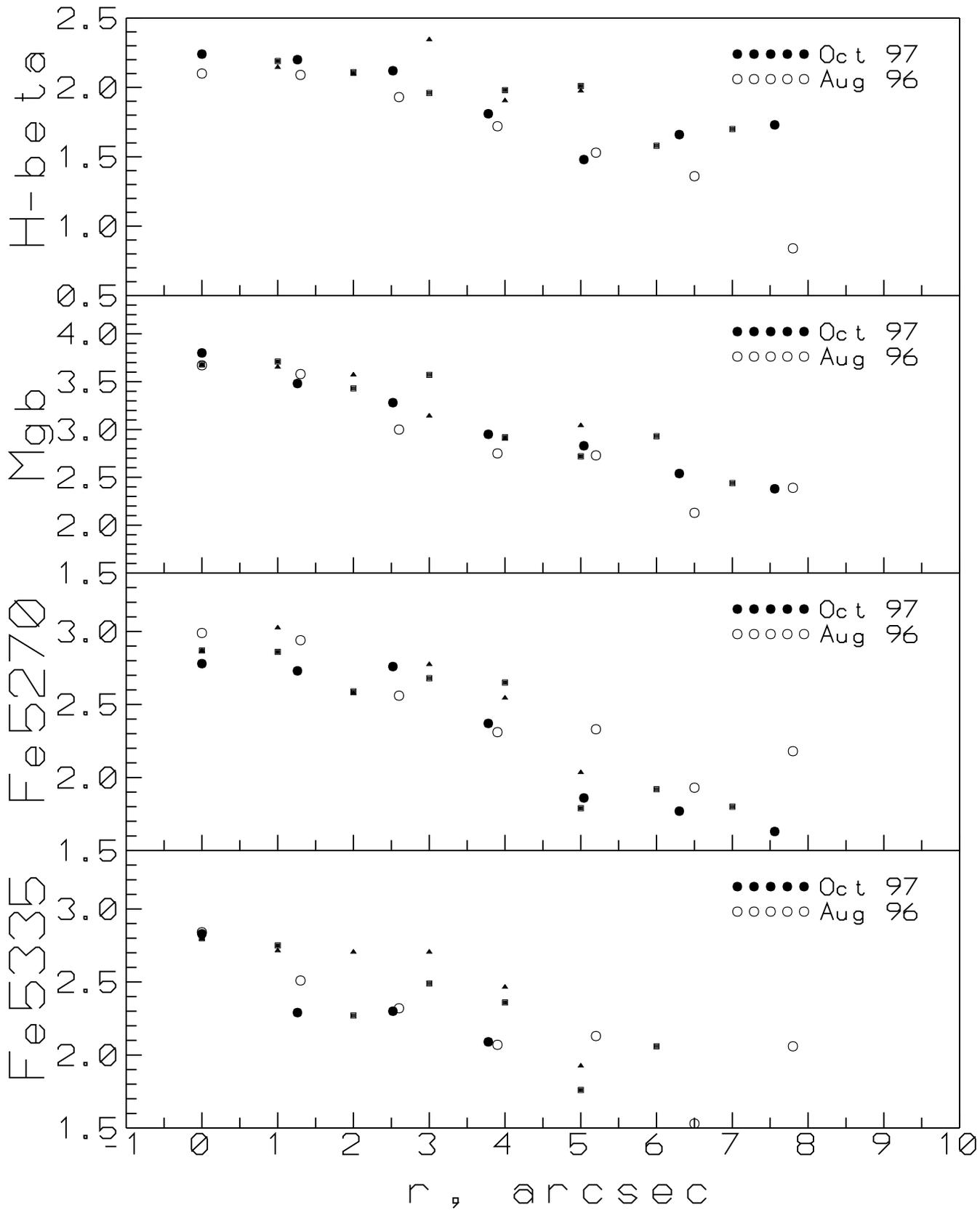
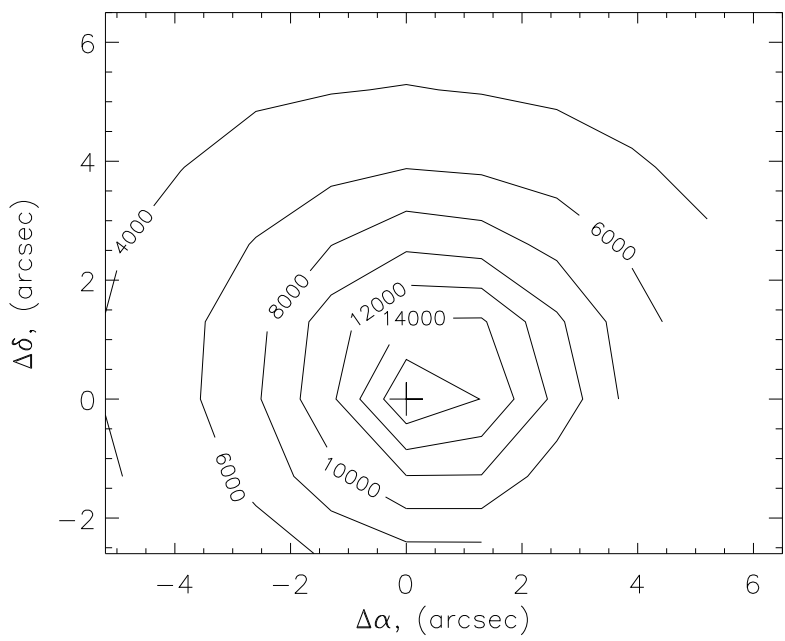
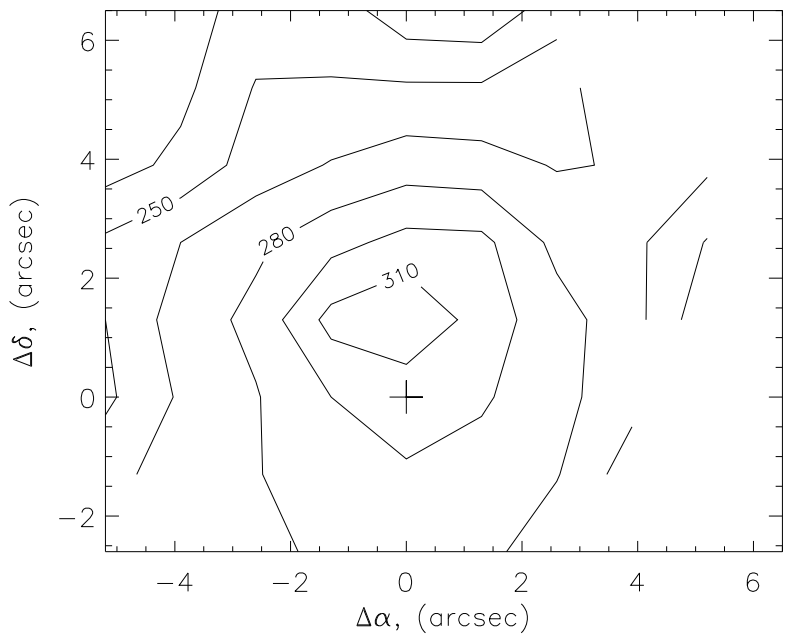
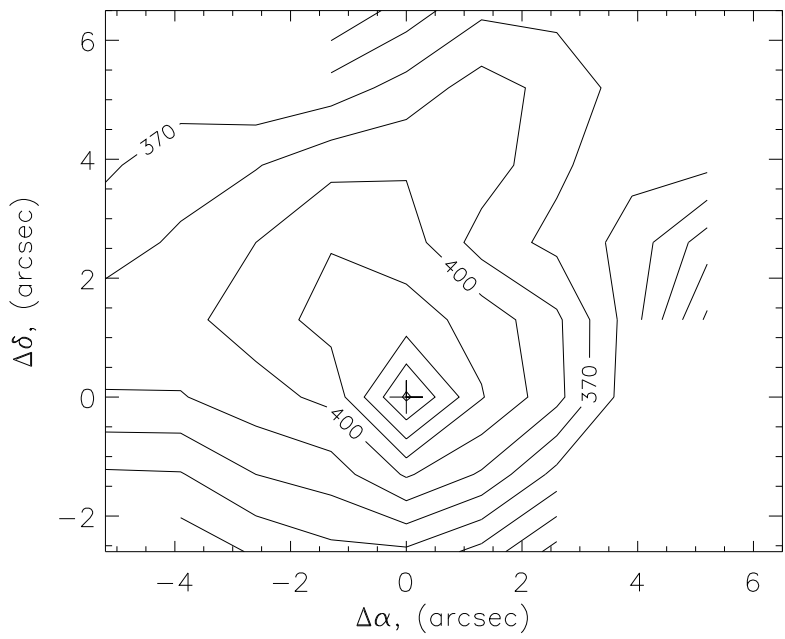
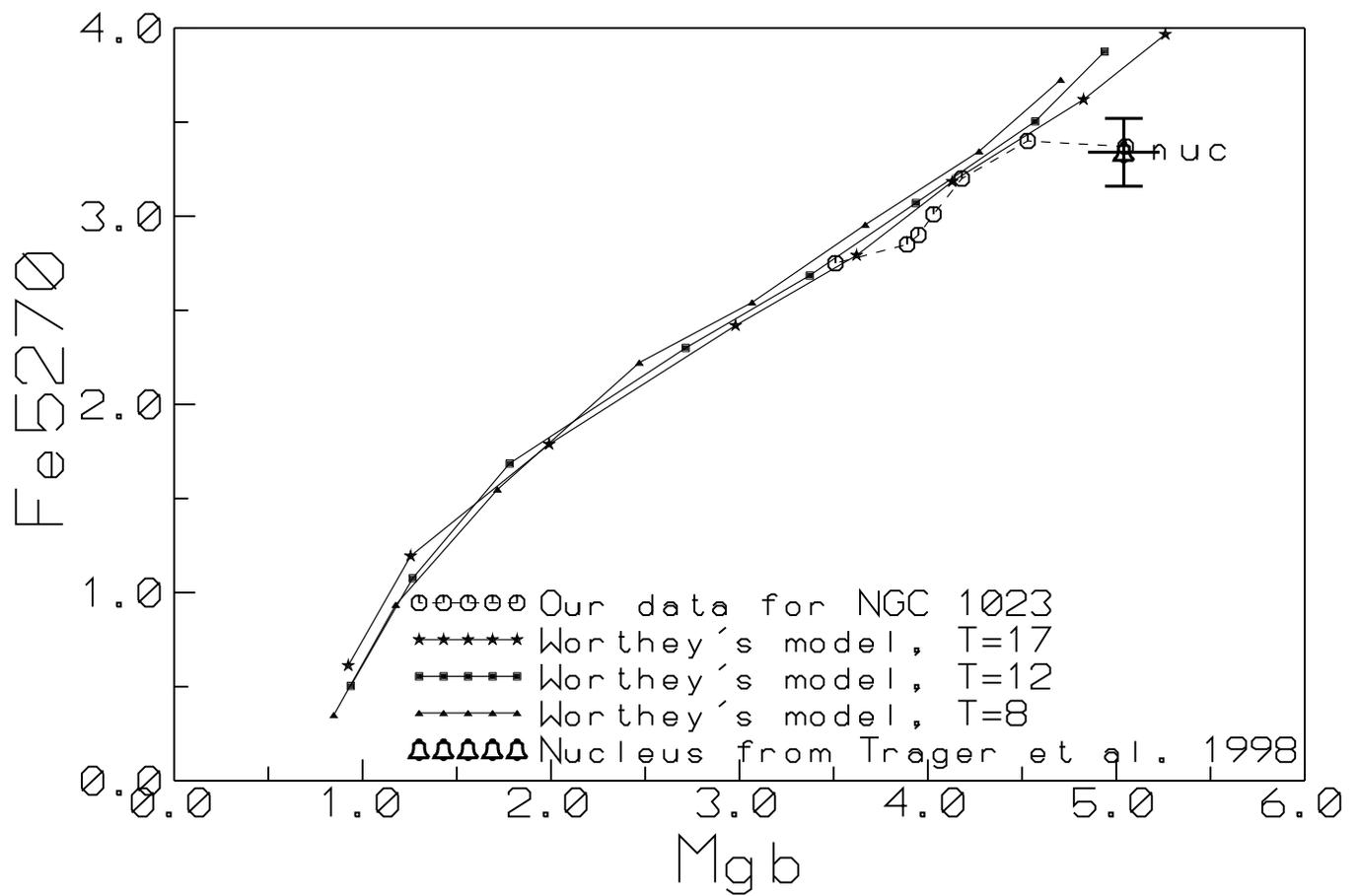
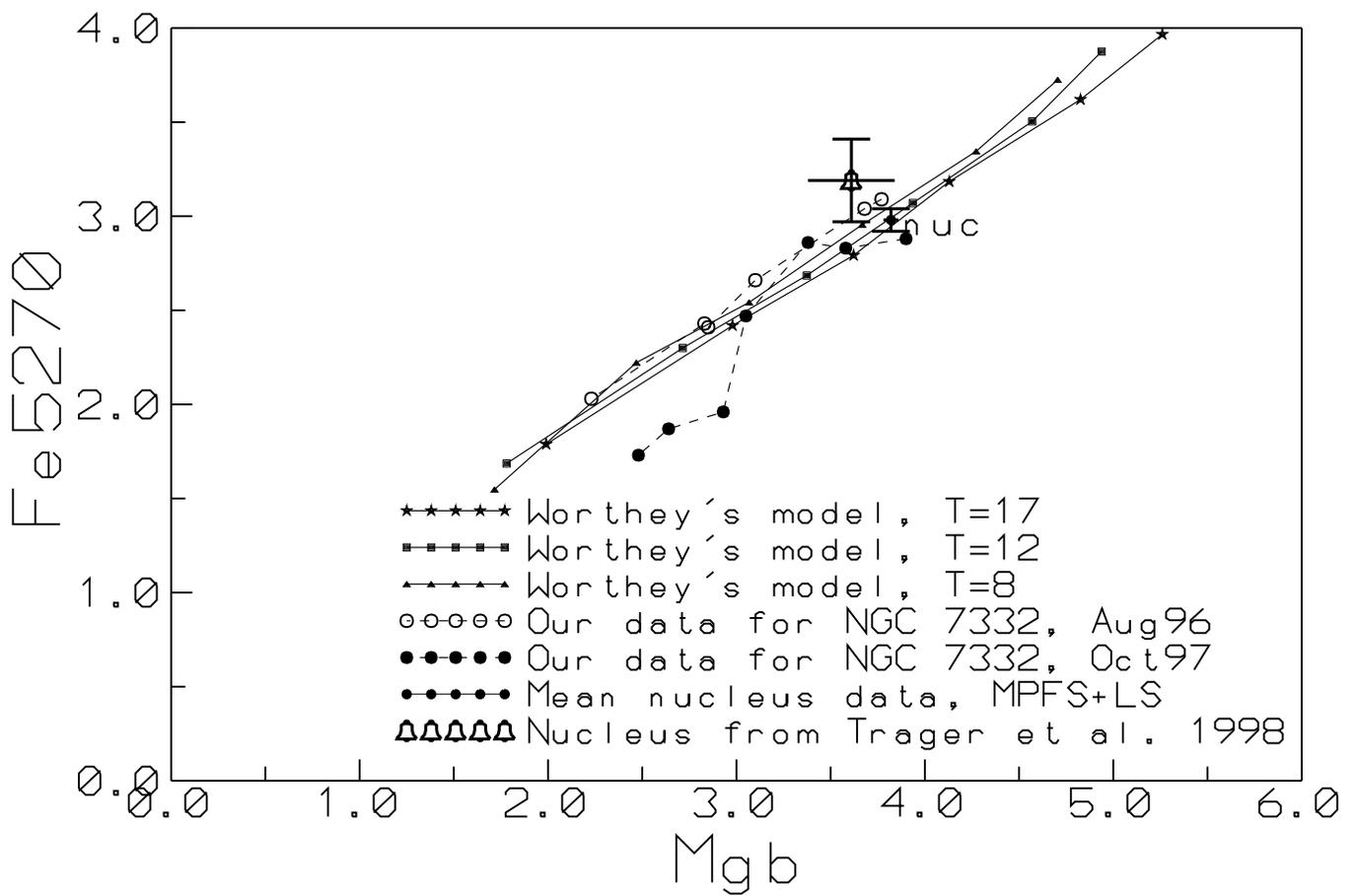
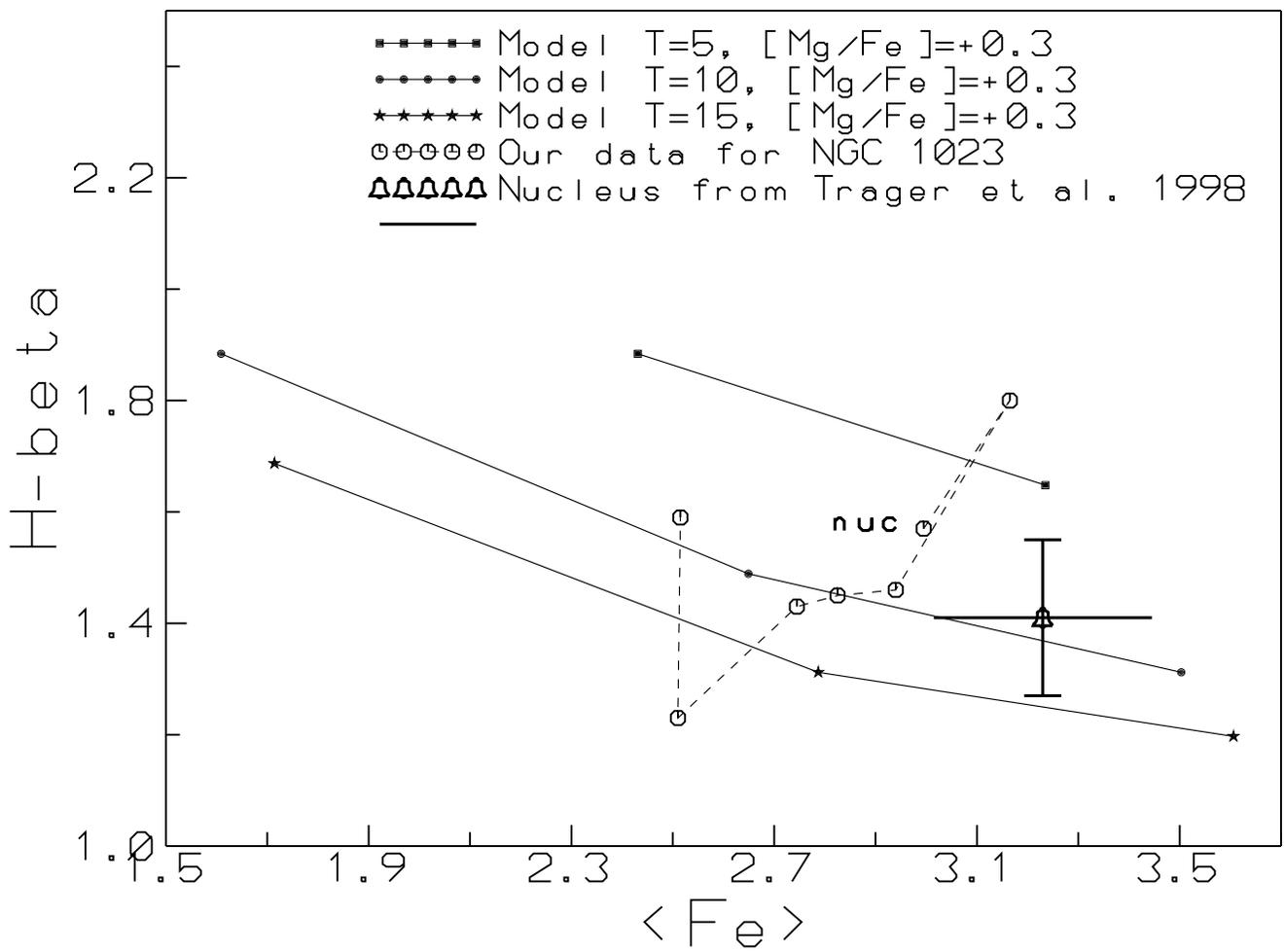
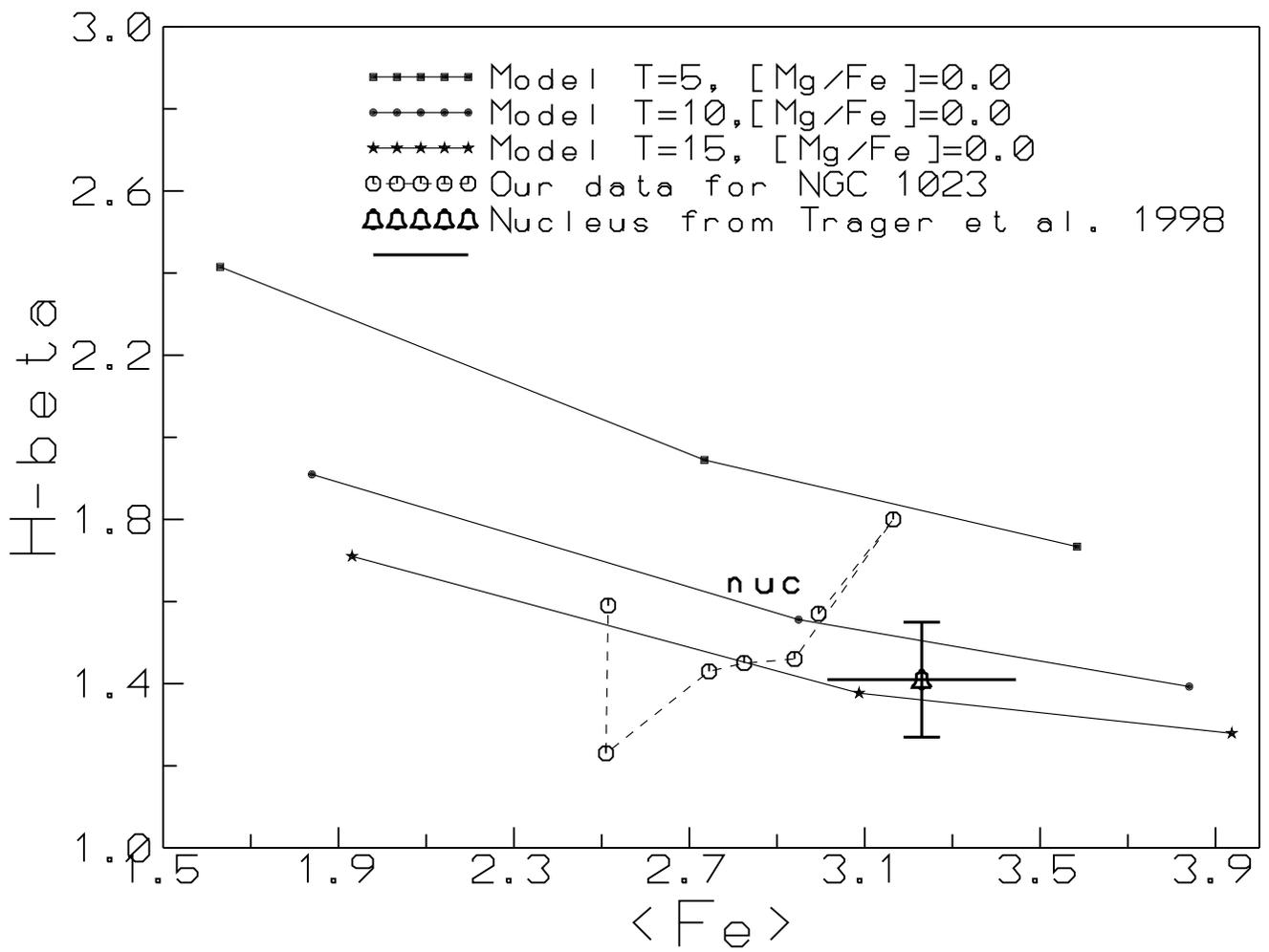
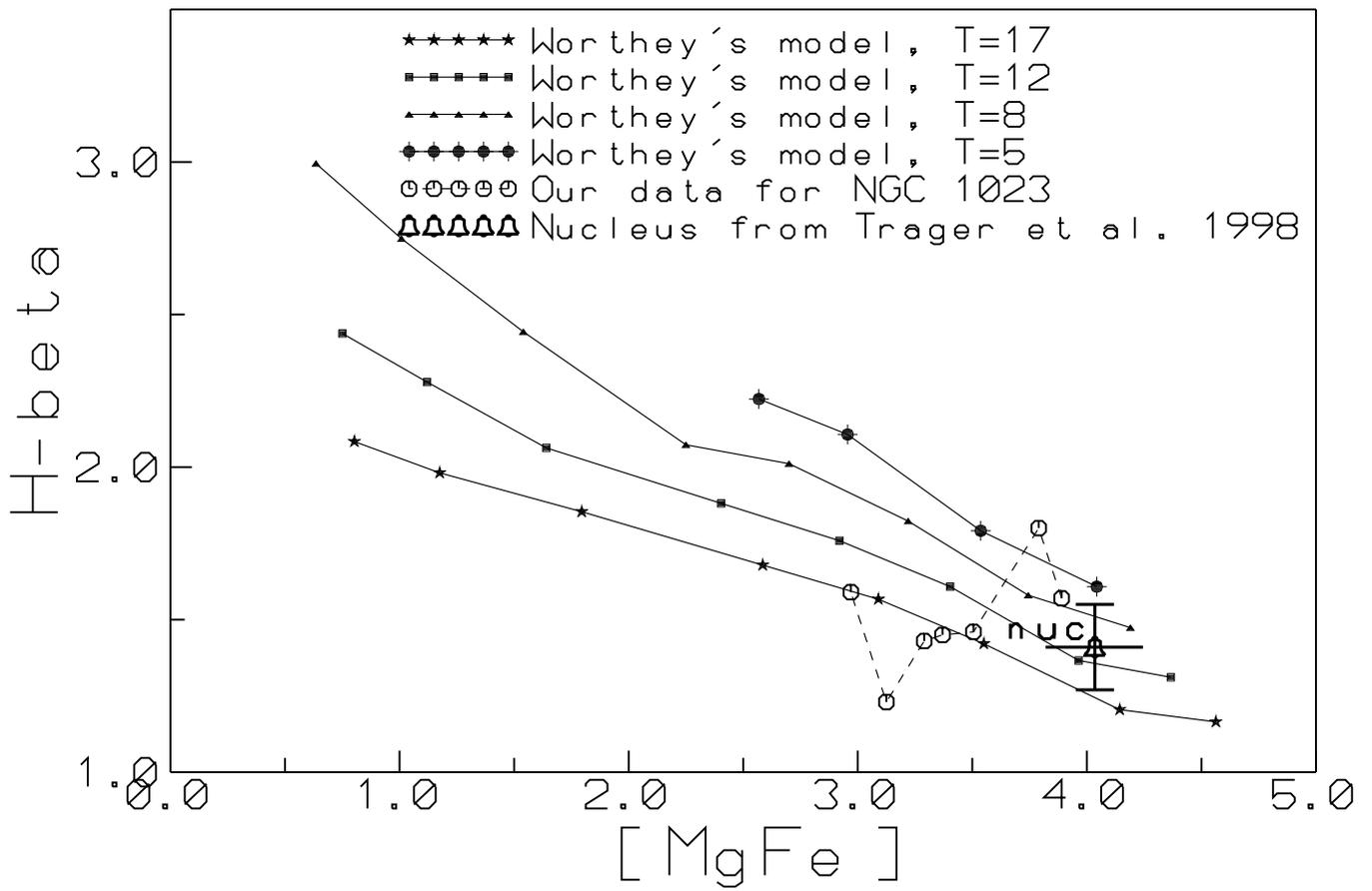
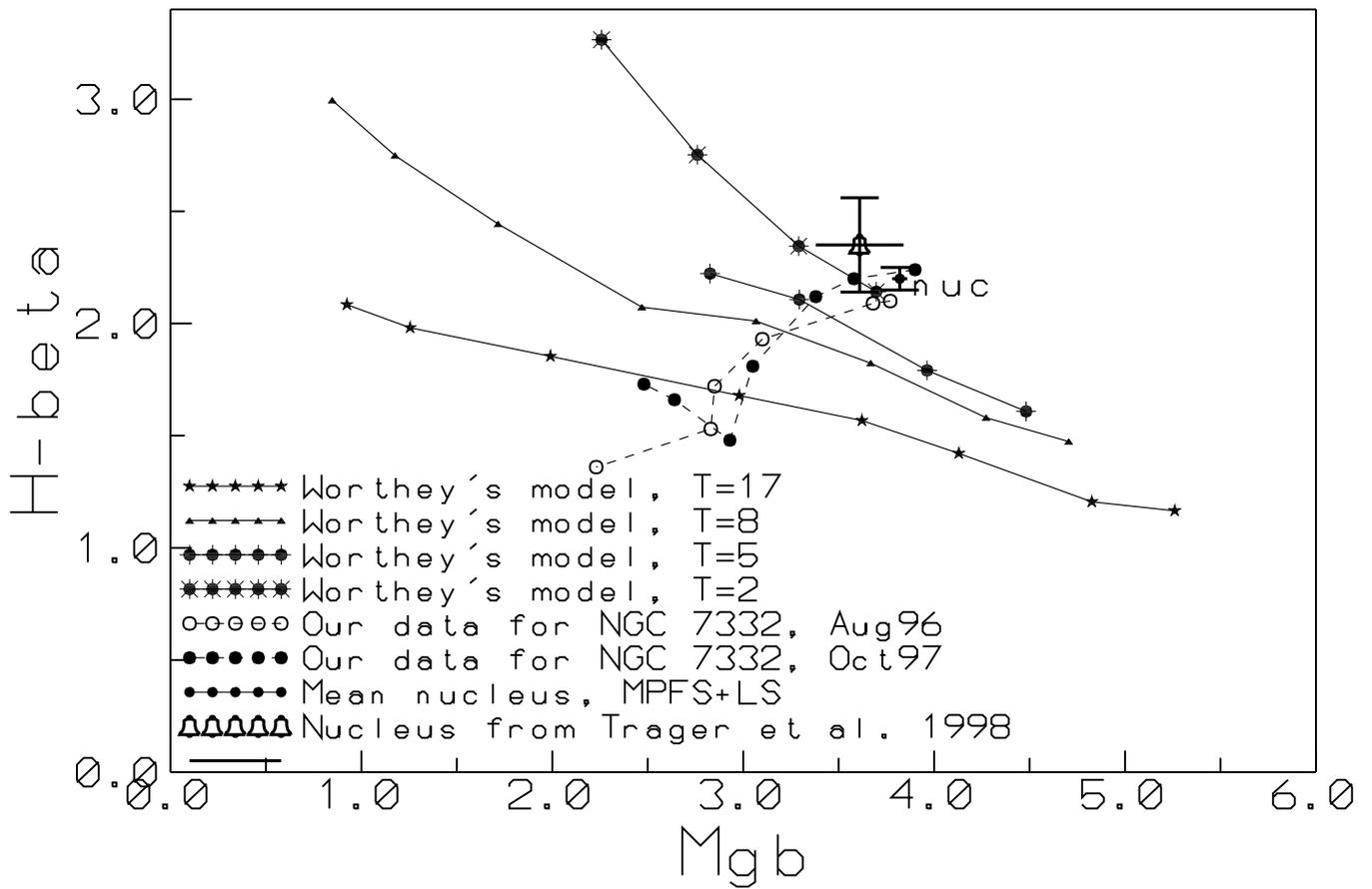
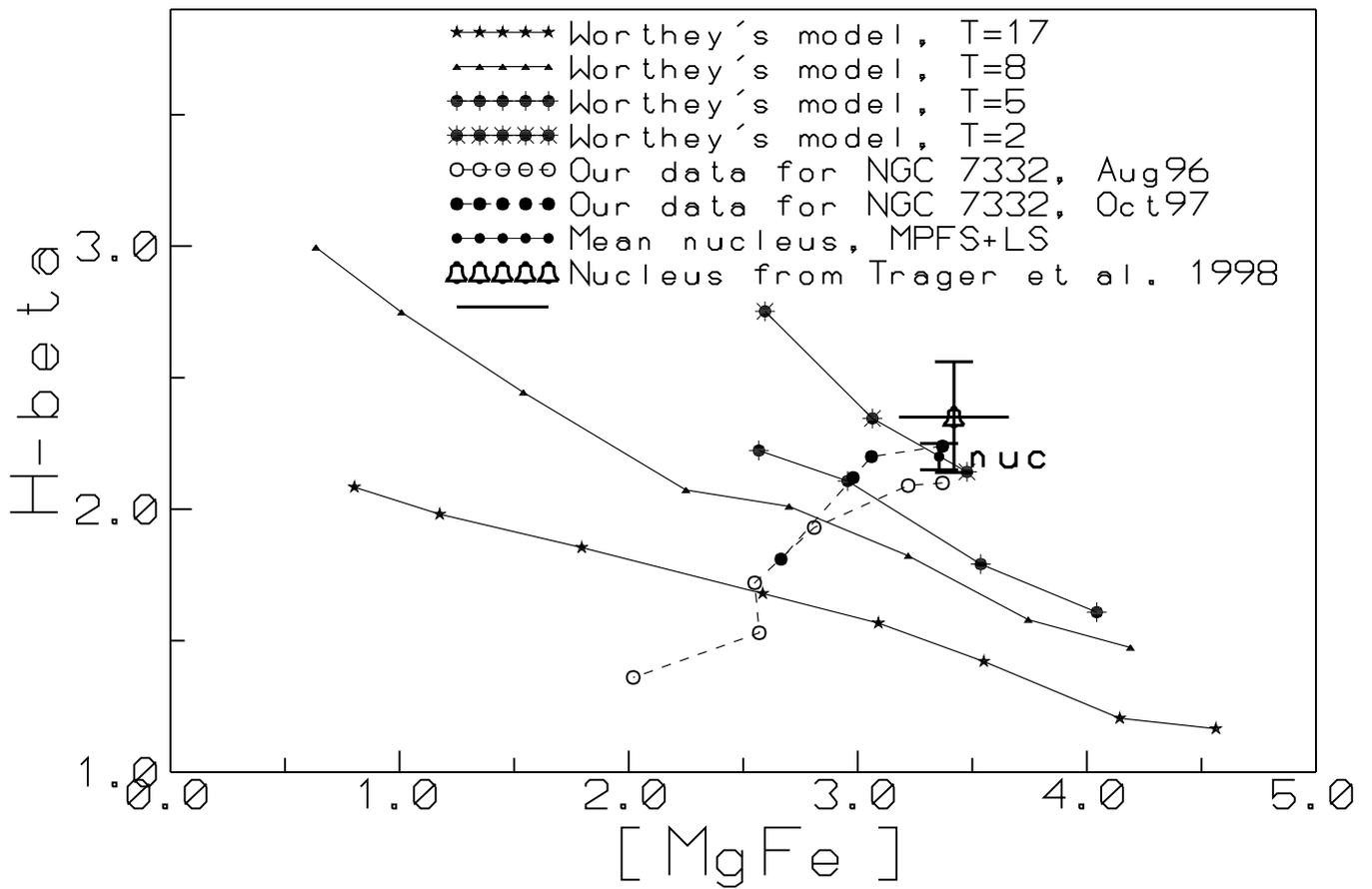
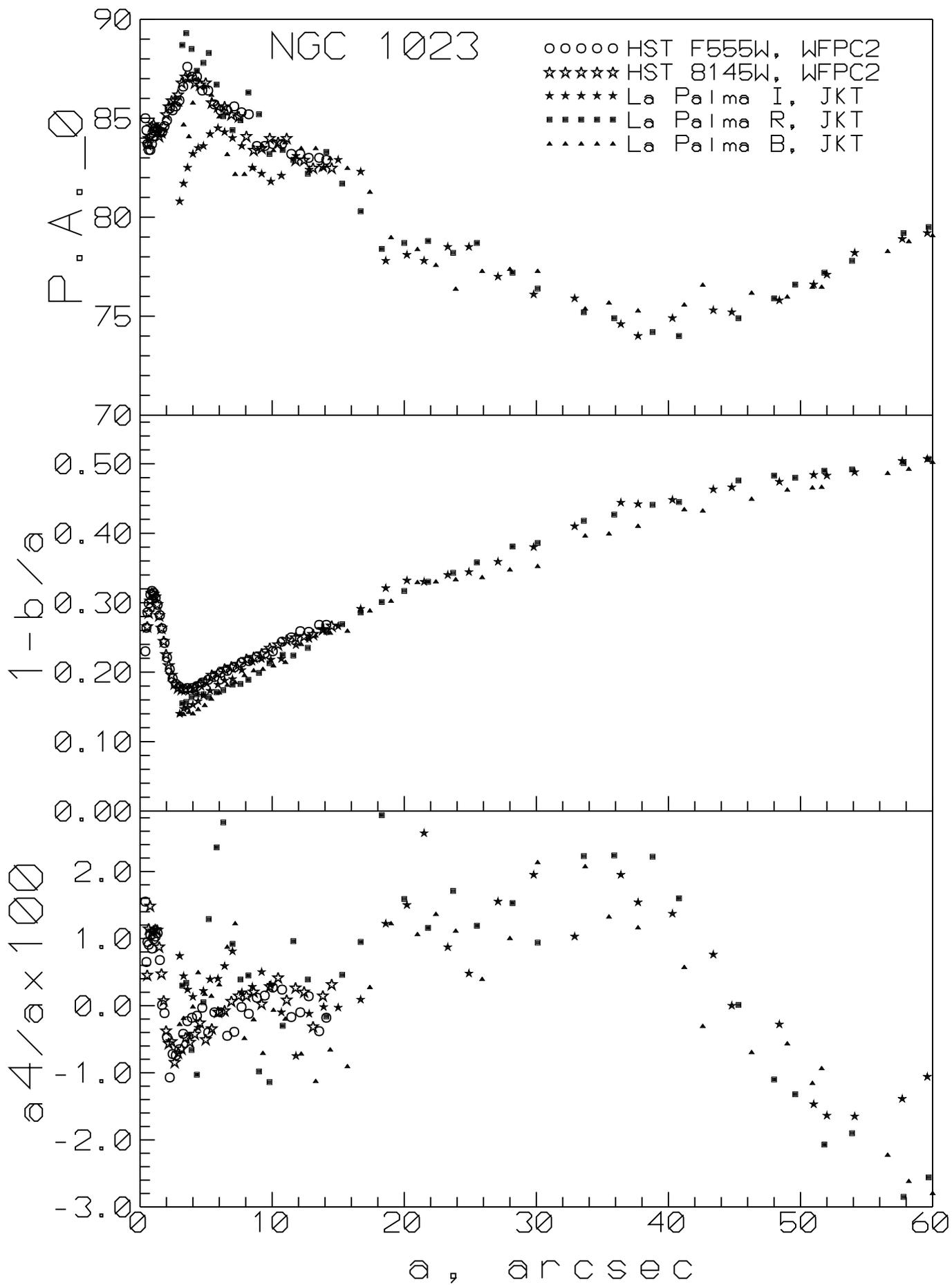
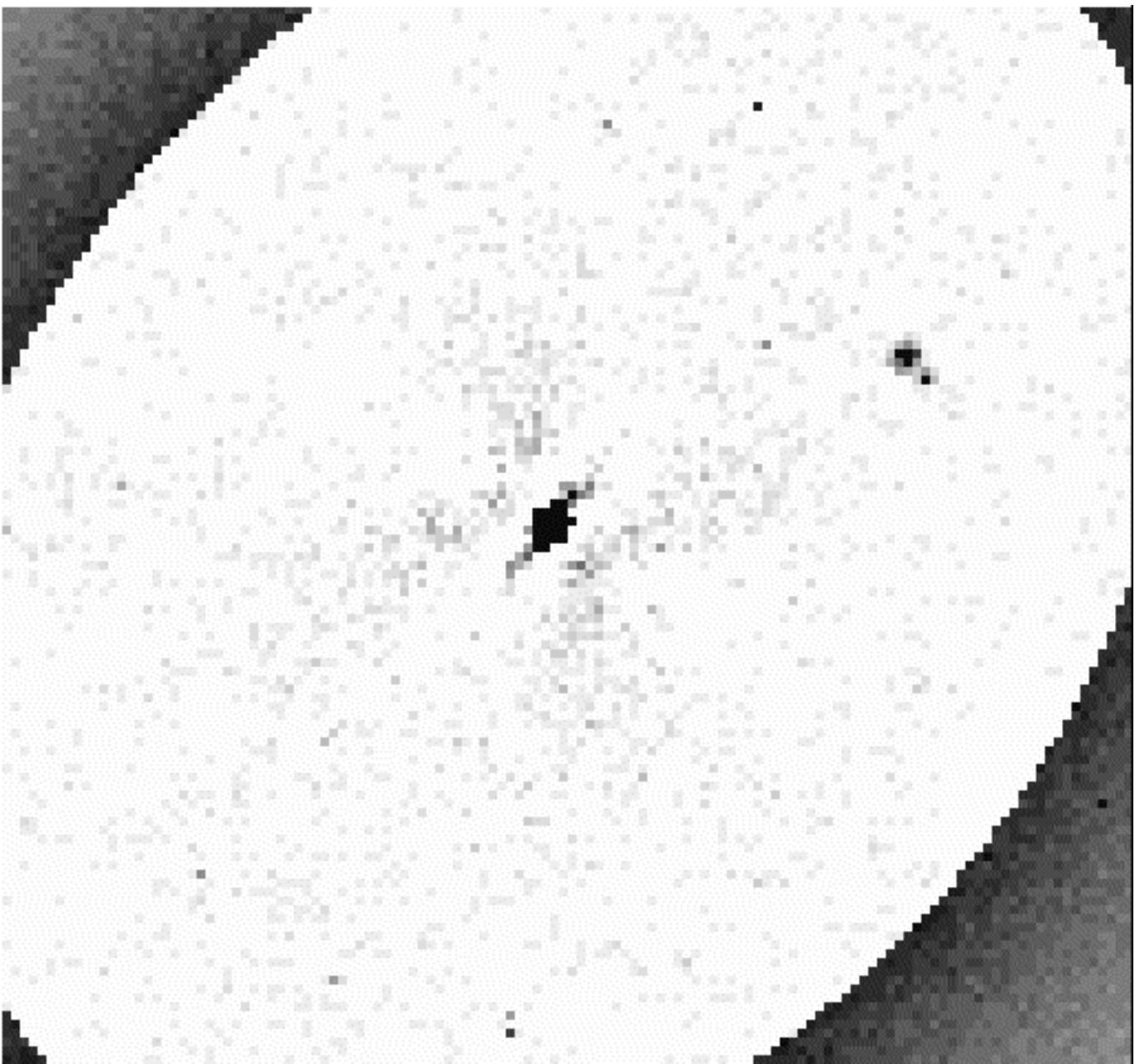
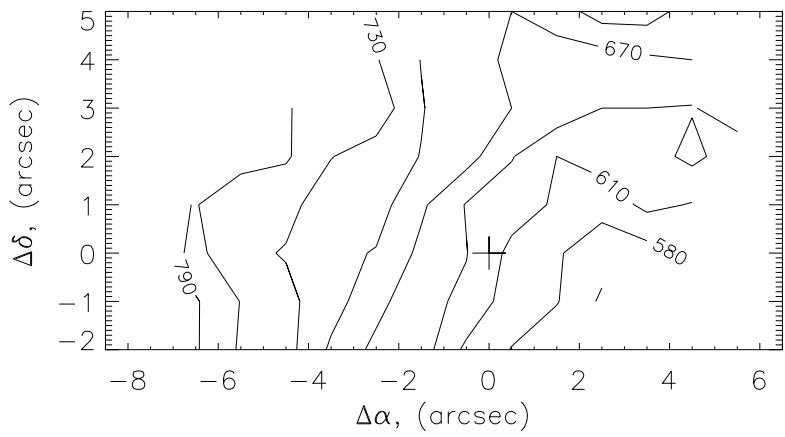
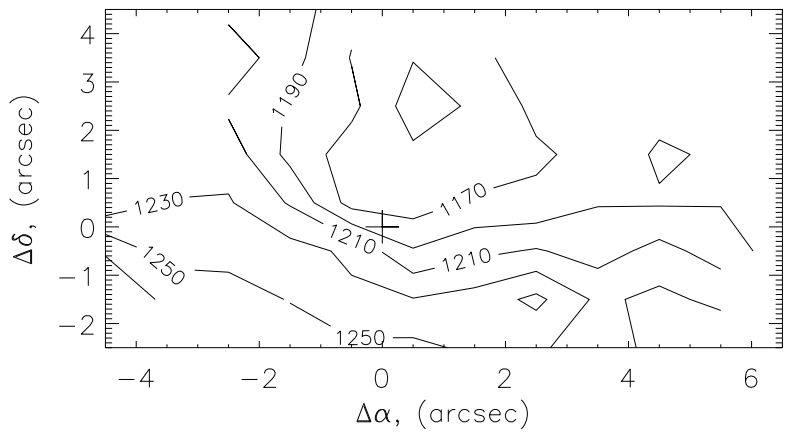
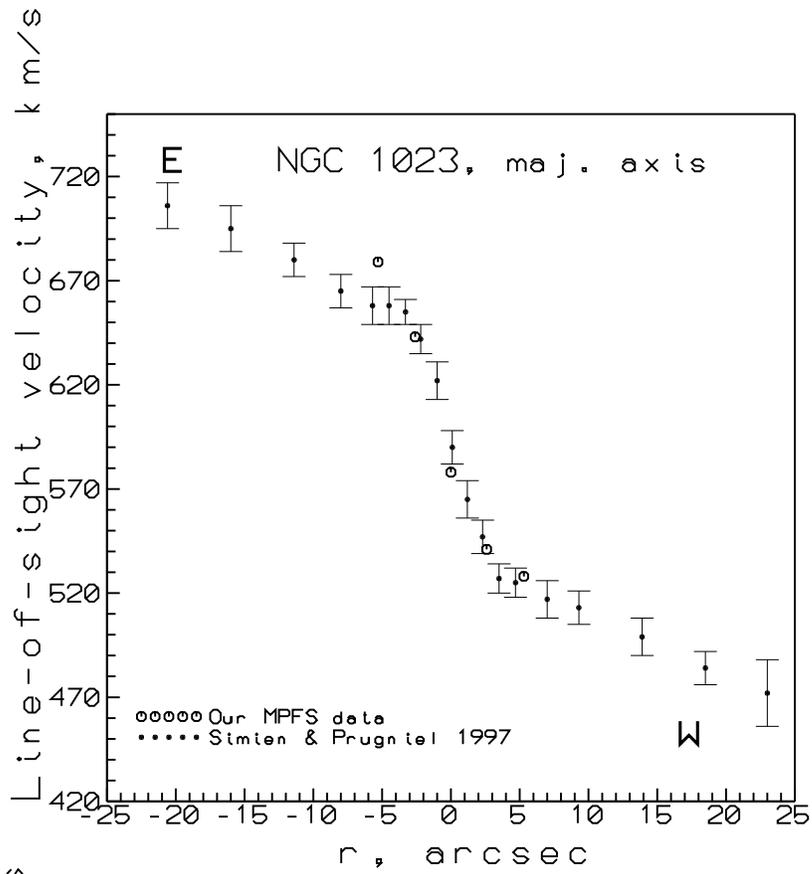
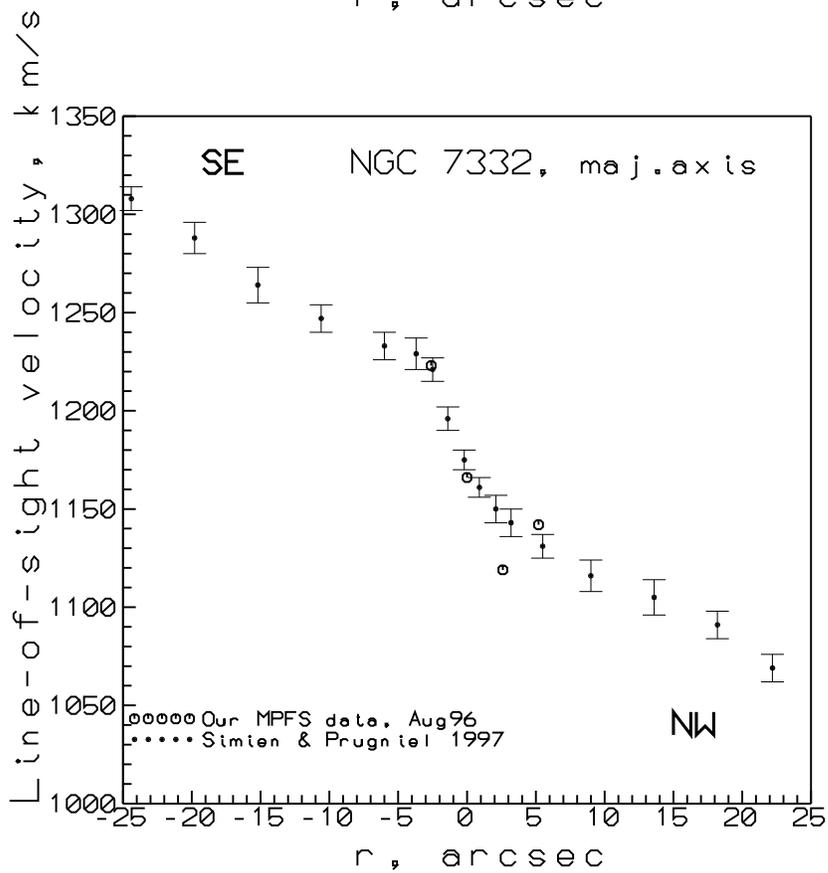
\end{document}